\pdfoutput=1

\documentclass{article}
\usepackage{spconf,amsmath,graphicx}
\usepackage{verbatim}
\usepackage{amsmath}
\usepackage{amssymb}        
\usepackage{xcolor}
\usepackage{empheq,float}
\usepackage{graphicx,caption,subcaption,balance}
\usepackage{epstopdf}

\input{mysymbol.sty}

\title{Forecasting Multi-dimensional Processes over graphs}
\name{Alberto Natali, Elvin Isufi and Geert Leus\thanks{This work was supported in parts by the KAUST-MIT-TUD consortium grant OSR-2015-Sensors-2700.}}
\address{Faculty of Electrical Engineering, Mathematics and Computer Science\\ Delft University of Technology, Delft, The Netherlands  \\
E-mails: \{a.natali; e.isufi-1; g.j.t.leus\}@tudelft.nl}
\begin{document}
\ninept
\maketitle
\begin{abstract}
The forecasting of multi-variate time processes through graph-based techniques has recently been addressed under the graph signal processing framework. However, problems in the representation and the processing arise when each time series carries a vector of quantities rather than a scalar one. To tackle this issue, we devise a new framework and propose new methodologies based on the graph vector autoregressive model. More explicitly, we leverage product graphs to model the high-dimensional graph data and develop multi-dimensional graph-based vector autoregressive models to forecast future trends with a number of parameters that is independent of the number of time series and a linear computational complexity.
%
Numerical results demonstrating the prediction of moving point clouds corroborate our findings.
\end{abstract}
\vspace{-3mm}
\section{Introduction}\vspace{-2mm}
\label{sec:intro}

Forecasting time processes is crucial in biology, finance, and position estimation, where historical information is exploited to predict future trends and subsequently take optimal decisions \cite{biology} \cite{finance} \cite{box}. When the process is multi-dimensional, forecasting methods should also account for the structure hidden in the data to improve accuracy. The canonical tool to exploit these relations is the vector autoregressive (VAR) recursion \cite{timeseries}. Conventional VAR models are nevertheless overparameterized; hence, requiring a large amount of training data. One way to reduce the number of parameters in VAR models is to express the process relations through a graph \cite{mei} \cite{isufi}. In that case, the VAR model parameter matrices are substituted with graph filters leading to a number of trainable parameters that is independent of the process dimension as well as to a lower implementation complexity. As such, the so-called graph VAR model in \cite{isufi} has led to a superior prediction accuracy compared to the classical VAR model.

The graph VAR models each time series as a scalar signal evolving over the nodes of a graph. However, it does not account for situations where multiple time series are related to a single node. In a position estimation problem, where each node is a point in the 3D space, we have three-dimensional processes evolving over each node. A similar problem is also encountered when predicting related quantities in weather monitoring sensor networks, e.g., temperature, humidity, air quality, and chemical concentrations. While the graph VAR model from \cite{isufi} can be adopted here to predict the evolution of each time series separately, it will not account for the additional hidden relations in the multi-dimensional process at each node.

To account for this additional information and improve the prediction accuracy, we devise a product graph VAR model that at one hand captures the inter-relations between the processes of different nodes, while on the other hand captures the intra-relations in the multi-dimensional process at a single node. Product graphs enjoy a wide popularity in modeling structured information in large structured data sets \cite{prodgraph}. They have been used to aid sampling strategies for graph data \cite{dragotti} \cite{ortiz} \cite{varmakov}, build graph wavelets on circulant graphs \cite{kotz}, represent a graph process as a time-invariant graph signal on a larger graph \cite{grassi} \cite{romero}, and to build graph neural networks \cite{gama-mimo}. In this work, we use product graphs to aid forecasting of multi-dimensional processes on graphs. One of the components in the product graph will model the inter-relations between nodes, while the other will capture the intra-relations between features of a node. We subsequently build on the concept of product graph filters \cite{bda} to put forth a general parametric VAR model that can forecast multi-dimensional processes with a number of parameters that is independent of the graph and feature dimensions. The latter is also extended to a novel approach that parameterizes also the type of product graph and allows learning an ad-hoc structure for forecasting purposes. Numerical results for position prediction of a $3$D point cloud corroborate the proposed method and its potential to improve accuracy compared with the single-feature graph models \cite{isufi}.

The rest of this paper proceeds as follows. Section~\ref{sec:format} lays down some preliminary material and the graph VAR model of \cite{isufi}. Section~\ref{sec:pagestyle} contains the proposed product graph VAR model. Section~\ref{sec:numerical-results} contains the numerical results, while Section~\ref{sec:majhead} concludes the paper.

\section{Graph VAR model}\vspace{-2mm}
\label{sec:format}

Consider an $N-$dimensional time series $\bbx_t \in \reals^N$, in which each entry $x_{t}(i)$ is a time-varying signal for a quantity of interest; $x_{t}(i)$ can be a temperature recording by a sensor $i$ at time $t$, while $\bbx_t$ collects the recordings of $N$ such sensors. We can model the temporal evolution of $\bbx_t$ through the vector autoregressive model:\vskip-1mm
\begin{equation}
\label{eq:var}
\bbx_{t}= - \sum_{p=1}^{P} \mathbf{A}_{p} \bbx_{t-p} + \bbvarepsilon_{t},
\end{equation}
which expresses the current value $\bbx_{t}$ as the linear combination of $P$ past realizations $ \bbx_{t-1}, \ldots, \bbx_{t-P}$. The $N \times N$ matrices $\bbA_{p}$ contain the $N^2$ parameters of this model and express the influence of the different entries of $\bbx_{t-p}$ into $\bbx_{t}$; vector ${\bbvarepsilon}_{t} \in \mathbb{R}^{N}$ collects the model error also labeled as the \textit{innovation} term \cite{timeseries}. Estimating the $PN^2$ parameters in \eqref{eq:var} is challenging, especially when $N$ is large. As such, parametric models for $\mathbf{A}_{p}$ are necessary and popular choices include factor models \cite{factor1} \cite{factor2} or low-rank data representations \cite{timeseries}.

When the relations between the different time series $x_t(i)$ in $\bbx_t$ can be captured by a network, we can exploit this structure to reduce the $PN^2$ parameters of \eqref{eq:var} in an efficient way that does not hurt prediction accuracy. To be more precise, let graph $\mathcal{G= (V,E}, \mathbf{S})$ model the network, where $\mathcal{V}=\{1, \ldots, N\}$ is the set of nodes (or vertices), $\mathcal{E} \subseteq \mathcal{V} \times \mathcal{V}$ is the set of edges, and $\bbS$ is an $N \times N$ matrix to represent this graph structure. We refer to the matrix $\bbS$ as the graph shift operator matrix \cite{shuman} ---e.g., the weighted adjacency matrix $\mathbf{W}$ (directed graphs) or the graph Laplacian $\mathbf{L}$ (undirected graphs)--- with non-zero entries $[\bbS]_{ij}$ only if $\left(i,j\right) \in \mathcal{E}$ or $i = j$. For a fixed $t$, $\bbx_{t}$ is a graph signal in which entry $x_t(i)$ resides on node $i$ \cite{shuman}.

We can process \emph{instantly} the graph signal $\bbx_{t}$ over the graph by using so-called graph filters \cite{moura}; by setting $\bbx_{t}$ as the filter input, we get the filtered signal:\vskip-.2cm
\begin{equation}
    \label{eq:filtering}
    \bby_{t} = \sum_{k=0}^{K} h_{k}\bbS^{k}\bbx_t := \mathbf{H}(\mathbf{\bbS})\bbx_{t},
\end{equation}
where the polynomial matrix $\mathbf{H}(\bbS)= \sum_{k=0}^{K} h_{k} \mathbf{S}^{k}$ is the graph filter with scalar coefficients $h_0, \ldots, h_K$. The expression in \eqref{eq:filtering} builds the instantaneous output entry $y_t(i)$ of $\bby_t$ at node $i$ as a linear combination of $K+1$ terms: the first term is the signal value $x_t(i)$ of node $i$; the other $K$ terms are contributions of signal values $x_t(j)$ from $K$-hop neighbors of node $i$. Since $\bbS$ respects the sparsity of the graph, the operation $\bbS\bbx_t$ aggregates at node $i$ the signal values of its one-hop neighbors with a complexity $\ccalO(|\ccalE|)$, where $|\ccalE|$ is the number of edges. Likewise, $\bbS^K\bbx_t = \bbS(\bbS^{K-1}\bbx_t)$ aggregates at node $i$ the signal values of its $K$-hop neighbors with a complexity  $\ccalO(K|\ccalE|)$, which is also the complexity of the operation in \eqref{eq:filtering}.

The graph filtering operation in \eqref{eq:filtering} represents a powerful way to instantly process $\bbx_t$ by accounting for the underlying structure. Graph filters have been used in \cite{isufi} to substitute the $P$ matrices $\bbA_p$ in \eqref{eq:var} with $P$ different graph filters for modeling $\bbx_t$ as:\vspace{-.1cm}
\begin{align}
\label{eq:g-var}
\begin{split}
    \bbx_{t} =-\sum_{p=1}^{P}\bbH_p(\bbS)\bbx_{t-p} + \bbvarepsilon_t =-\sum_{p=1}^{P} \sum_{k=0}^{K} h_{kp} \bbS^{k} \bbx_{t-p}+{\bbvarepsilon}_{t}.
    \end{split}
\end{align}
In other words, each of the past realizations $\bbx_{t-1}, \ldots, \bbx_{t-P}$ is treated as a different graph signal, which is further processed with a different graph filter $\bbH_p(\bbS)$. As such, the process value $x_t(i)$ at node $i$ depends now on the process values with a temporal window of $P$ and a graph window of $K$, i.e., $x_t(i)$ depends on the past $P$ values of the nodes that are within $K$ hops from node $i$. The graph VAR (G-VAR) model in \eqref{eq:g-var} inherits from the graph filters a number of parameters that is $P(K+1)$ and an implementation complexity of $\ccalO(PK|\ccalE|)$. Both these quantities are orders smaller, and independent on the graph process dimensions, than the respective $PN^2$ and $\ccalO(PN^2)$ of the classical VAR in \eqref{eq:var}; hence, requiring less training data to estimate the model parameters.

Despite the above benefits, the G-VAR model cannot handle cases where on each node we have a vector of feature values rather than a scalar $x_t(i)$. Treating each feature separately and predicting its evolution with \eqref{eq:g-var} is certainly an option but it will not account for the dependencies between features. To account efficiently for these dependencies and improve the prediction accuracy, we introduce next the notion of multi-dimensional graph processes and leverage product graphs to develop generalized G-VAR models for forecasting the temporal evolution of such processes.


\section{Product Graph VAR Model}\vspace{-2mm}
\label{sec:pagestyle}

\begin{figure*}%
\centering
\begin{subfigure}{0.33\textwidth}
\centering
\includegraphics[width=0.55\textwidth]{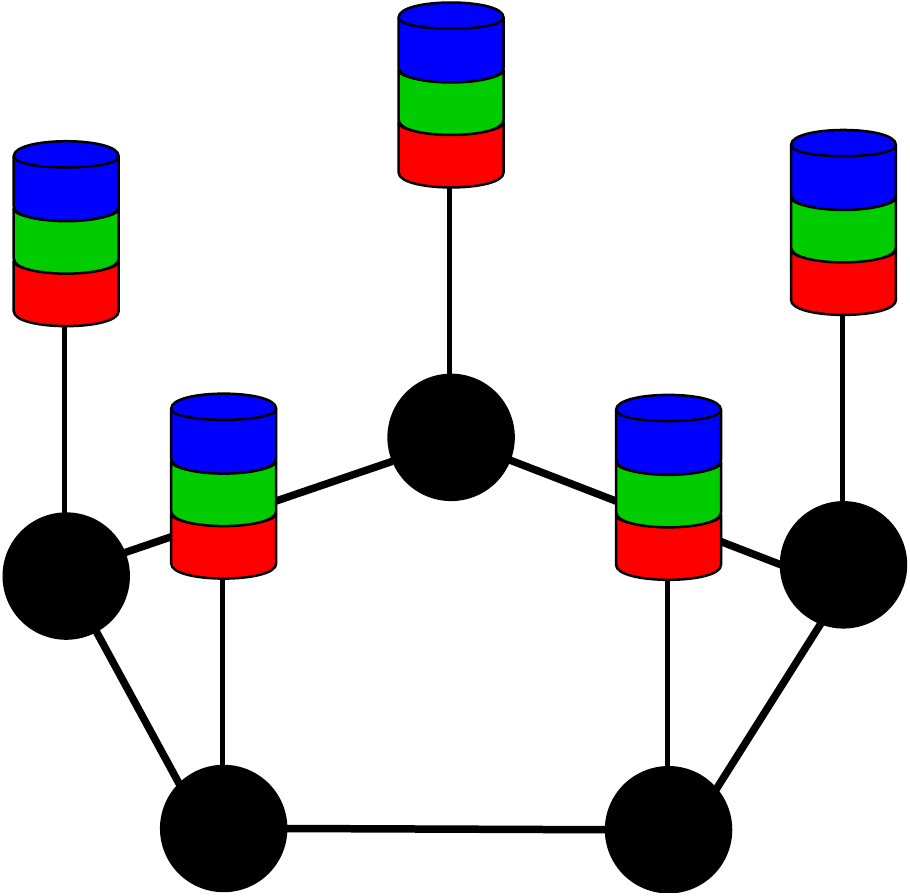}%
\caption{Three-dimensional graph signal.}%
\label{sfig_mdgs}%
\end{subfigure}%
\begin{subfigure}{0.33\textwidth}
\centering
\includegraphics[width=0.65\textwidth]{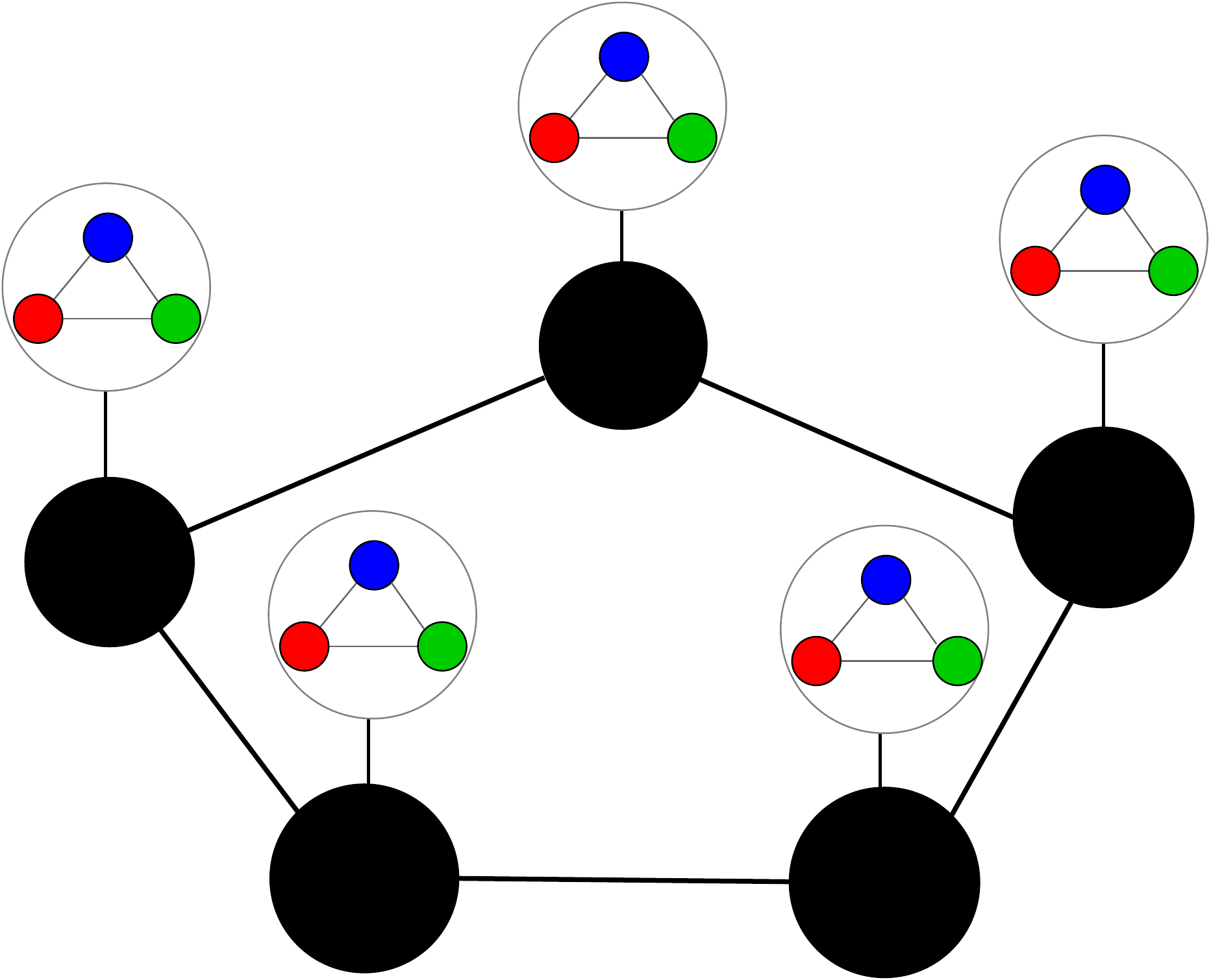}%
\caption{Feature graph over each node.}%
\label{sfig_featG}%
\end{subfigure}%
\begin{subfigure}{0.33\textwidth}
\centering
\includegraphics[width=1\textwidth]{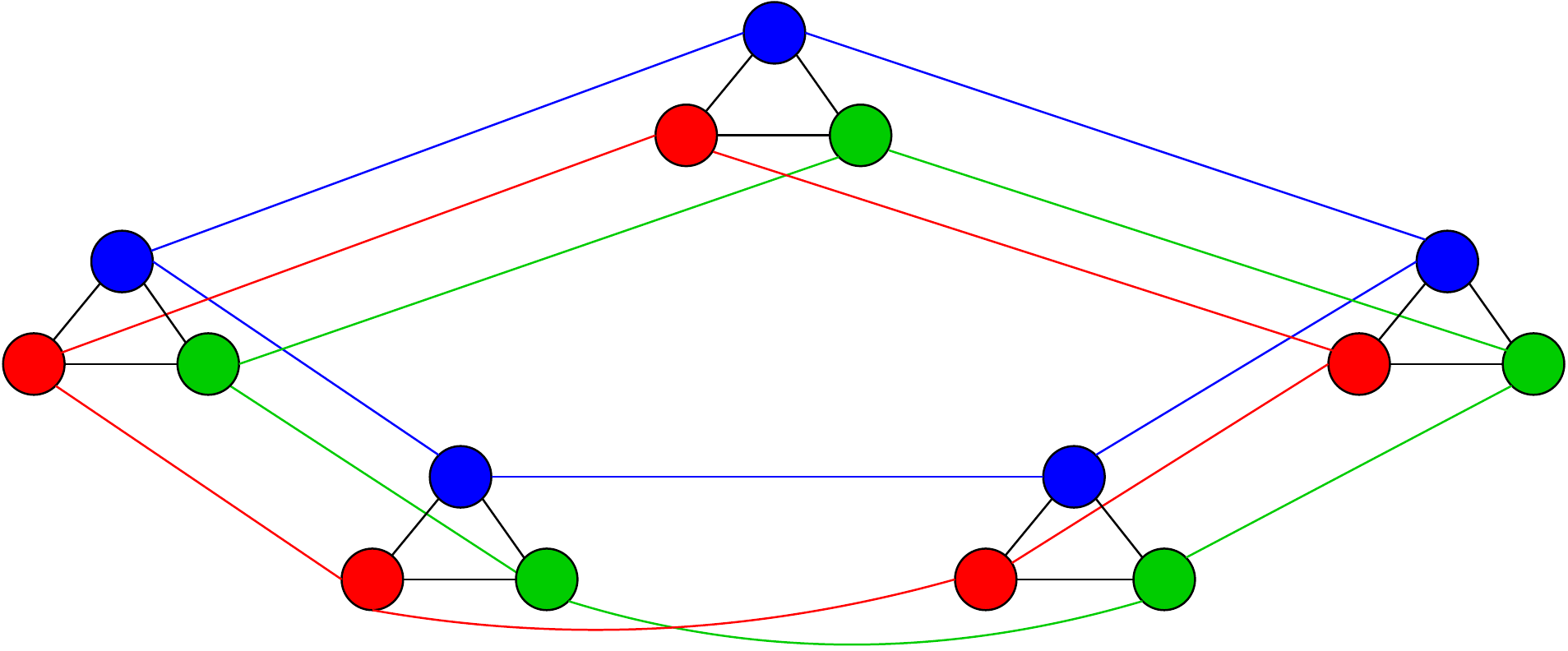}%
\vspace{.7cm}
\caption{Multi-dimensional graph signal on product graph.}%
\label{sfig_pg}%
\end{subfigure}%
\caption{Multi-dimensional graph signal represented through a product graph. (a) A three-dimensional graph signal. Each color represents a different feature. We can see the three-dimensional graph signal in two ways: $(i)$ each node has a node signal of dimension three composed by the three different colors; $(ii)$ the collection of each feature (e.g., blue) for all nodes forms a feature signal of dimension five (i.e., the number of nodes). (b) A feature graph formed between the features of each node. Each colored circle (red-green-blue) is now a node in the feature graph and the edges connecting them are the edges of the feature graph. (c) Cartesian product graph illustration of the results in (b). The feature graph in (b) influences the intra-connectivities between the features of each node (shown in black); the underlying graph in (a) influences the inter-connectivities between the features of different nodes (shown by the respective feature color to ease visualisation).}\label{fig:vary_n1}\vspace{-5mm}
\end{figure*}

When on top of each node we have a vector of $F$ feature values rather than a scalar, we talk about a \textit{multi-dimensional} graph signal. We store the feature values of node $i \in \ccalV$ in the $F \times 1$ vector $\bbx_t(i)$ and refer to it as the \emph{node signal} for node $i$. The overall \emph{F-dimensional graph signal} at time $t$ is the $NF \times 1$ vector $\bbx_t = [\bbx_t^{\top}(1), \ldots, \bbx_t^{\top}(N)]^\top$, i.e., the vector that concatenates all node signals\footnote{Note that the  $F$-dimensional graph signal is denoted with the same symbol $\bbx_t$, as the one-dimensional graph signal in Section~\ref{sec:format}. We chose this slight abuse of notation to construct a generalized G-VAR model that follows a similar recursion as in \eqref{eq:g-var}.}. Likewise, we can also consider the $f$th feature of all nodes $\{x_t^{(f)}(i)\}_{i \in \ccalV}$ and store them in the $N \times 1$ vector $\bbx_t^{(f)} = [x_t^{(f)}(1), \ldots, x_t^{(f)}(N)]^\top$, which we refer to as the \emph{feature signal}. With the terminology of Section~\ref{sec:format}, each feature signal is a graph signal and the $F$-dimensional graph signal is a collection of $F$ (one-dimensional) graph signals. Fig.~\ref{sfig_mdgs} illustrates a three-dimensional graph signal.

Our goal now is to forecast the temporal evolution of the $F-$dimensional graph process $\bbx_t$. We want to develop a graph VAR model as in \eqref{eq:g-var} to forecast now the evolution of an $NF \times 1$ vector $\bbx_t$. A naive way doing this is to ignore the underlying graph structure $\ccalG$ and build an alternative graph (using topology identification based on a certain metric \cite{connecting}) of $NF$ nodes and use the G-VAR model \eqref{eq:g-var} to forecast the process. This strategy has two main disadvantages: first, it destroys the known underlying structure present between nodes (e.g., friendships in a social network) and builds a feature-based similarity graph; second, it leads to a shift operator of larger dimensions without any further structure, hence with a larger storage and computational complexity. To tackle both these issues, we use the concept of product graphs to propose a product graph VAR model for forecasting $F$-dimensional graph processes.

\subsection{Product Graph Representation}\vspace{-1mm}

We assume that the features of each node signal $\bbx _t(i)\in \reals^F$ have a hidden relation which we represent with a \emph{feature graph} $\mathcal{G}_{\mathcal{F}} = \left( \mathcal{V_{F}, E_{F}}, \mathbf{S}_{\mathcal{F}}\right)$, where each vertex in $\ccalV_\ccalF = \{1, \ldots, F\}$ is one feature, $\ccalE_{\ccalF} \subseteq \mathcal{V}_\ccalF \times \mathcal{V}_\ccalF$ is the edge set connecting different features, and $\mathbf{S}_{\mathcal{F}}$ is the $F \times F$ graph shift operator matrix of $\mathcal{G}_{\mathcal{F}}$. Observe that all nodes have the same feature graph ${\mathcal G}_{\mathcal F}$. A simple way to build $\ccalG_\ccalF$ is through a feature similarity metric, e.g., Pearson correlation or Gaussian weighting kernels \cite{connecting}. In other words, we consider each feature graph $\ccalG_\ccalF$ now residing on the nodes $\ccalV$ of the underlying graph $\ccalG$. Fig.~\ref{sfig_featG} illustrates possible feature graphs between the features of each node.

While the graph $\ccalG_\ccalF$ influences the intra-connections between the features of a node $i$, the graph $\ccalG$ influences the inter-connections between different nodes $i, j \in \ccalV$. In particular, if nodes $i$ and $j$ share an edge in $\ccalG$, this edge should be reflected in the inter-connections between the features of the vectors $\bbx_t(i)$ and $\bbx_t(j)$ and, vice-versa, if $i$ and $j$ are not connected in $\ccalG$, there should not be inter-connectivities between the features of the vectors $\bbx_t(i)$ and $\bbx_t(j)$. We can formally capture both the intra- and  inter-connectivities between features with product graphs. The product graph of the underlying graph $\ccalG$ and the feature graph $\ccalG_\ccalF$ is a new graph:
\begin{equation}
\label{eq:product-graph}
    \mathcal{G}_{\diamond}=  \mathcal{G}  \diamond \mathcal{G}_{\mathcal{F}} =\left(\mathcal{V}_{\diamond}, \mathcal{E}_{\diamond} ,\mathbf{S}_{\diamond}\right),
\end{equation}
with node set $\ccalV_\diamond$ of cardinality $|\mathcal{V}_{\diamond}|=NF$, and where the edge set $\ccalE_\diamond \subseteq \mathcal{V}_\diamond \times \mathcal{V}_\diamond$ and the $NF \times NF$ graph shift operator $\bbS_\diamond$ are dictated by the type of product graph. Popular product graphs are the Kronecker, Cartesian, and strong product \cite{prodgraph}. For the Cartesian product, for instance, the graph shift operator is $\mathbf{S}_{ \diamond}=\mathbf{S} \otimes \mathbf{I}_{F}+\mathbf{I}_{N} \otimes \mathbf{S}_{\mathcal{F}}$ with $\bbI_F$ ($\bbI_N$) the $F\times F$ ($N \times N$) identity matrix. The edge set cardinality for this graph is $|\ccalE_\diamond| = F|\ccalE| + N|\ccalE_\ccalF|$. Fig.~\ref{sfig_pg} illustrates the Cartesian product graph for the three-dimensional graph signal example. 

The $F$-dimensional graph signal $\bbx_t$ on $\ccalG$ is now a (one-dimensional) graph signal on $\ccalG_\diamond$. Therefore, we can develop a G-VAR model as in \eqref{eq:g-var} w.r.t. the product graph shift operator $\bbS_\diamond$. We will refer to the latter as a product graph VAR model, while we will also develop a more generalized version that goes beyond the direct application of \eqref{eq:g-var} to a product graph. The benefits of this product graph strategy are that the storage complexity of $\bbS_\diamond$ is that of storing $\bbS$ and $\bbS_\ccalF$ separately and the computational complexity is of the combined sparsity orders of $\bbS$ and $\bbS_\ccalF$; both in general lower than building a graph for the $NF \times 1$ signal $\bbx_t$.

\textbf{Product Graph VAR.} As discussed above, we thus model the temporal evolution of the $F$-dimensional graph process $\bbx_t \in \reals^{NF}$ with the product graph VAR (PG-VAR) model:
\begin{equation}
    \label{eq:pgp-var}
     \bbx_{t}= - \sum_{p=1}^{P} \sum_{k=0}^{K} h_{kp}\bbS_{\diamond}^{k}  \bbx_{t-p} + {\bbvarepsilon}_{t}.
\end{equation}
Model \eqref{eq:pgp-var} expresses, through $\bbS_\diamond$, the influence that the past $F$-dimensional graph process realizations $\bbx_{t-1}, \ldots, \bbx_{t-P}$ have on each node signal at time $t$, i.e., $\bbx_t(i)$ for $i \in \ccalV$. The order $K$ implies now that the $f$th feature $x_t^{(f)}(i)$ in $\bbx_t(i)$ depends on all other feature realizations $x_{t-p}^{(g)}(j)$ for $g = 1, \ldots, F$, $j \in{\ccalV}$, and $p = 1, \ldots, P$, within a $K$-hop neighborhood in the product graph $\ccalG_\diamond$. As such the PG-VAR model forecasts future values of $\bbx_t$ by accounting on the intra- and inter-connectivities between the different node features. Besides storage and computational benefits, model \eqref{eq:pgp-var} has also a number of parameters $P(K+1)$ that is independent on $F$ and $N$; hence, requiring little training data to estimate them. For $\ccalG_\diamond$ being the Cartesian graph, the implementation complexity of \eqref{eq:pgp-var} is $\ccalO\big(PK(F|\ccalE| + N|\ccalE_\ccalF|)\big)$ which is comparable with the cost $\ccalO(PKF|\ccalE|)$ of $F$ independent G-VAR models \eqref{eq:g-var}, since $N$ and $|\ccalE|$ (as well as $F$ and $|\ccalE_\ccalF|$) are often of the same order. But the PG-VAR model captures now additional intra-relations; hence, has the potential to improve the prediction accuracy.

\textbf{Generalized Product Graph VAR.} The performance of the PG-VAR heavily depends on the considered type of product graph. However, it is not clear which is the most suited product graph for a task at hand. Here, we by-pass this issue by jointly estimating the model parameters and the type of product graph. For an underlying graph $\ccalG = (\ccalV, \ccalE, \bbS)$ and a feature graph $\ccalG_\ccalF = (\ccalV_\ccalF, \ccalE_\ccalF, \bbS_\ccalF )$, any product graph can be expressed as:
\begin{align}
\label{eq:general-PG}
    \mathbf{S}_{\diamond}= \sum_{i = 0}^1 \sum_{j = 0}^1 s_{ij}  \left(\mathbf{S}^{i} \otimes \mathbf{S}_{\ccalF}^{j}   \right),
\end{align}
where ``$\otimes$" denotes the Kronecker product and $s_{ij} \in \{ 0,1\}$. A graph filter $\bbH(\bbS_\diamond)$ of order $K$ over the product graph $\ccalG_\diamond$ has thus the form:
\begin{align}
\label{eq:pg-filter}
\begin{gathered}
   \bbH(\bbS_\diamond)= \sum_{k=0}^{K} h_{k} \mathbf{S}_{\diamond}^{k} =\sum_{k = 0}^K h_{k} \left( \sum_{i,j = 0}^1 s_{ij}  \left(\mathbf{S}^{i} \otimes \mathbf{S}_{\ccalF}^{j} \right) \right) ^{k},
\end{gathered} 
\end{align}
where we see that the coefficients $s_{ij}$ that define the type of product graph play a role in the filtering behavior. The filter in \eqref{eq:pg-filter} can be seen as a special case of the more general product graph filter:
\begin{equation}
\label{eq:PG-maybe}
    \bbH(\bbS_\diamond) =\sum_{k = 0}^K\sum_{l = 0}^L {h}_{kl} \left(\mathbf{S}^{k} \otimes \mathbf{S}_{\ccalF}^{l} \right),
\end{equation}
which is now defined by the general parameters $h_{kl}$. The novel aspect of the general filter in \eqref{eq:PG-maybe} is in the way the shift operators $\bbS$ and $\bbS_\ccalF$ are combined within the filter. We can therefore use \eqref{eq:PG-maybe} to extend \eqref{eq:pgp-var} towards a general  PG-VAR  model for the $F$-dimensional graph process $\bbx_t$ as:
\begin{equation}
\label{eq:gpgp-var}
    \bbx_{t}=-\sum_{p=1}^{P} \sum_{k = 0}^K\sum_{l = 0}^L h_{lkp} \left(\mathbf{S}^{k} \otimes \mathbf{S}_{\ccalF}^{l} \right) \bbx_{t-p} + {\bbvarepsilon}_{t} .
\end{equation}
At a slight increase in the number of parameters to estimate, i.e., $PKL$, compared with model \eqref{eq:pgp-var}, the generalized PG-VAR model parameterizes now also the type of product graph. As such it exploits ad-hoc the intra- and inter-connectivities in the historical data of the $F$-dimensional process $\bbx_t$.

\begin{figure*}%
\centering
\begin{subfigure}{0.45\textwidth}
\centering
\includegraphics[width=\textwidth,trim={.1cm .1cm 0 0},clip]{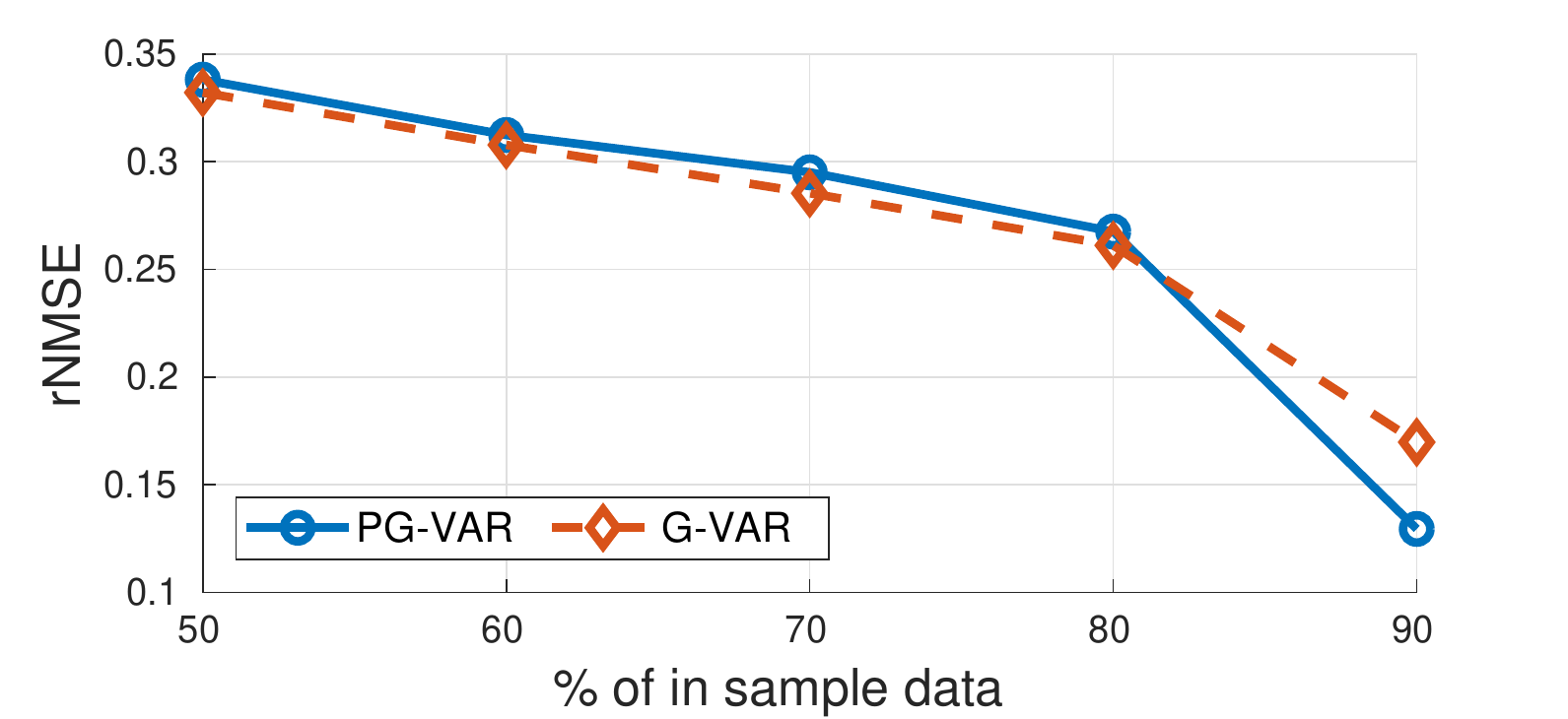}%
\caption{}%
\label{fig:avg}%
\end{subfigure}%
\begin{subfigure}{0.5\textwidth}
\centering
\includegraphics[width=.75\textwidth,trim={1cm 1.1cm 0 0},clip]{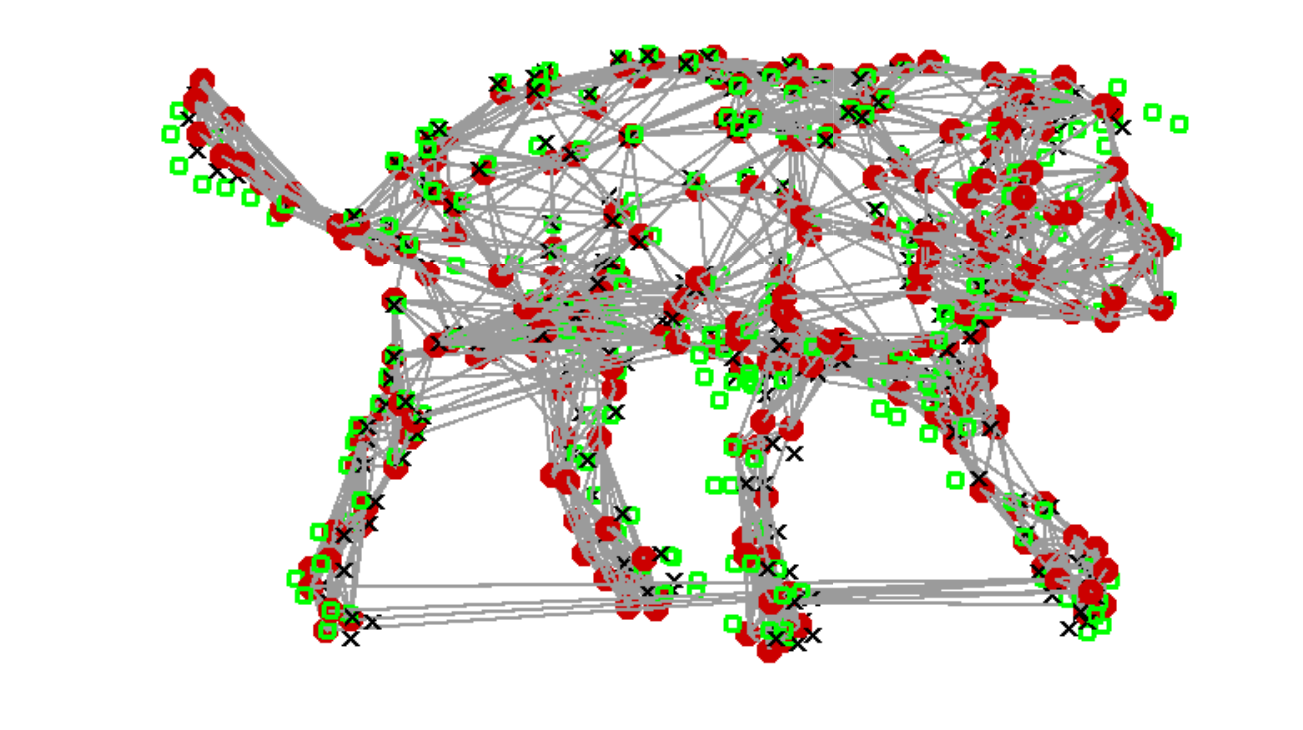}%
\caption{}%
\label{fig:dog}%
\end{subfigure}%
\caption{Test rNMSE for the one-step ahead prediction of the two algorithms considering the entire multi-dimensional graph process versus the percentage of in-sample data. (c) One-step-ahead estimated position of the walking dog for a test sample. The red points correspond to the ground-truth position, while the black crosses and the green circles correspond to the PG-VAR and G-VAR predictions, respectively.}\label{fig:vary_n2}\vspace{-5mm}
\end{figure*}

\subsection{Parameter estimation}\vspace{-2mm}

One of the main benefits of models \eqref{eq:pgp-var} and \eqref{eq:gpgp-var} is that their respective parameters can be readily estimated following the procedure for the G-VAR developed in \cite{isufi}. Let $\bbH_p(\bbS_\diamond)$ for $p = 1, \ldots, P$ be a collection of $P$ graph filters that can represent either the forms \eqref{eq:pgp-var} or  \eqref{eq:gpgp-var}. Given the historical data $\{\bbx_{1}, \bbx_{2}, \ldots, \bbx_{t-1}\}$ of the $F$-dimensional graph process, the optimal predictor for $\bbx_t$ is given by the conditional expectation:
\begin{align}
\label{eq:optimal}
\begin{gathered}
   {\tbx}_{t} = \mathbb{E}\left[\bbx_{t} |\left\{\bbx_{1}, \ldots, \bbx_{t-1}\right\}\right]=  - \sum_{p=1}^{P} \mathbf{H}_{p}(\bbS_\diamond ) \bbx_{t-p},
\end{gathered}
\end{align}
since the innovation term $\bbvarepsilon_t$ has a zero mean. The mean square error $\text{MSE} \big(\bbH_p(\bbS_\diamond)\big) = \mbE\left[\left\| \bbx_{t} - \tbx_{t} \right\|_{2}^{2}\right]$ for the optimal predictor \eqref{eq:optimal} is:
\begin{align}\label{eqn_mse}
    \begin{gathered}
        \text{MSE} \big(\bbH_p(\bbS_\diamond)\big)=\text{tr} \bigg( \bbR_0 + \sum_{p=1}^{P} \bbH_p(\bbS_\diamond )_{p}\bbR_p + \sum_{p=1}^{P} \bbR_p \mathbf{H}_{p}^\top(\bbS_\diamond ) \\
        + \sum_{p_{1}=1}^{P} \sum_{p_{2}=1}^{P}\bbH_{p_{1}}(\bbS_\diamond ) \bbR_{p_2 - p_1} \mathbf{H}_{p_{2}}^\top (\bbS_\diamond) \bigg),
    \end{gathered}
\end{align}
where $\bbR_p$ is the autocorrelation matrix of process $\bbx_t$ at temporal lag $p$, i.e., $\bbR_p = \mbE\big[\bbx_t\bbx_{t-p}^\top   \big]$. We then find the parameters for model \eqref{eq:optimal} as those that minimize the MSE. For the PG-VAR model, this implies solving the optimization problem:
\begin{equation}
\begin{aligned}
& \underset{\{h_{kp}\}}{\text{minimize}}
& & \text{MSE} \big(\bbH_p(\bbS_\diamond)\big) \\
& \text{subject to}
& & \bbH_p(\bbS_\diamond) = \sum_{k = 0}^K h_{kp}\bbS_\diamond,
\end{aligned}
\end{equation}
where the MSE expression is given in \eqref{eqn_mse}. Likewise, for the general PG-VAR model, the framework is similar, with the only difference in the filter expression $\bbH_p(\bbS_\diamond)$ and in the  optimization variables $h_{klp}$.

\section{Numerical Results}\vspace{-2mm}
\label{sec:numerical-results}

We consider the task of predicting the position of a walking dog mesh \cite{dog}. The mesh has $N = 251$ spatial points over $T = 59$ time instants. For each spatial point, we have the three $(x,y,z)$ coordinates that change over time. We treated each point as the nodes of a ten nearest neighbor ($10$NN) graph built from the coordinates in the first time instant. The node coordinates represent the $F = 3$ features and our goal is to predict the one-step value. As in \cite{isufi}, we pre-processed the data to render them with zero mean and unitary maximum value. We compared the PG-VAR model in \eqref{eq:pgp-var} with the G-VAR model in \eqref{eq:g-var}, where for this dataset the latter has yielded a superior performance compared to the standard VAR in \eqref{eq:var} and other graph-based techniques \cite{isufi}. We did not analyze the generaliz PG-VAR in \eqref{eq:gpgp-var} since the number of time instants is relatively low for the increased number of parameters; we will analyze this model on bigger data sets and present these observations in the extended version of this work.

\textbf{Experimental setup.} We considered the Cartesian product graph to model the relations between the underlying $10$NN graph $\ccalG$ and a fully connected feature graph $\ccalG_\ccalF$ of $F = 3$ nodes. The resulting product graph $\ccalG_\diamond$ has $753$ nodes connecting different $(x,y,z)$ coordinates. We followed \cite{isufi} and split the data across the temporal dimension into in-sample and out-of-sample data. We analyzed different percentages of this split ranging from $50\%$ to $90\%$ in-sample data. The in-sample data were used to estimate the model parameters and the out-of-sample data to test the performance. The in-sample data have been further divided into $70\%$ training and $30\%$ validation sets. We found the filter coefficients by fitting the model with different parameters $P$ and $K$ in the training set and assessing their performance in the validation set \cite{cross-validation}. Then, we refitted the model with the best performing tuple $(P,K)$ into the whole in-sample set. We evaluated the performance of the different algorithms with the root normalized MSE (rNMSE) defined as:
\begin{equation}
\mathrm{rNMSE}=\sqrt{\frac{\sum_{t=1}^{\tau}\left\|\tbx_{t}-\bbx_{t}\right\|_{2}^{2}}{\sum_{t=1}^{\tau}\left\|\bbx_{t}\right\|_{2}^{2}}},
\end{equation}
where $\tbx_t$ is the predicted value at time $t$ and $\bbx_t$ is the true one. For the G-VAR model in \eqref{eq:g-var}, we predicted each coordinate separately and report the average prediction error over all coordinates.

\textbf{Results.} Fig.~\ref{fig:avg} shows the rNMSE of the two methods for different percentages of in-sample data when considering the prediction of an entire multi-dimensional graph signal in the test set. Despite the considered scenario has a low number of training data and does not show high correlation between the different features, our method yields at least a comparable accuracy w.r.t. \eqref{eq:g-var}. However, when more training samples are available, the PG-VAR yields an improved performance. In addition, Fig.~\ref{fig:dog} depicts the true mesh position and the two estimates at a random (test) time instant, where we can observe how the proposed model matches well the real dog position.

\section{Conclusion}\vspace{-2mm}
\label{sec:majhead}

We proposed a product graph-based model to forecast future values of multi-dimensional graph processes, i.e., processes which at each node have multiple time-varying features. The proposed approach builds first a feature graph between the different node features and exploits then the underlying structure between nodes to link the different features on the different nodes with product graphs. Subsequently, we incorporated into the VAR models the product graph structure to forecast the multi-dimensional process with a number of parameters that is independent of the dimensions of the graphs and therefore has a low implementation complexity. Further, we also devised a general forecasting with product graphs, which learns directly from the data also the adequate type of product graph for the task at hand. As future works, we will corroborate our methods also on larger data sets,  execute numerical tests on the general PG-VAR model and  work on its possible  extension(s).

\vfill\pagebreak


\end{document}